\documentclass{article}
\usepackage{spconf,amsmath,graphicx}
\usepackage{subfig}
\usepackage{amsfonts}
\usepackage{multirow}
\usepackage{makecell}
\usepackage{mathrsfs}
\usepackage[ruled]{algorithm2e}

\PassOptionsToPackage{hyphens}{url}\usepackage{hyperref}

\title{ELECTROENCEPHALOGRAM SENSOR DATA COMPRESSION USING AN ASYMMETRICAL SPARSE AUTOENCODER WITH A discrete cosine transform LAYER}
\name{Xin Zhu$^*$ \quad Hongyi Pan$^\dagger$  \quad Shuaiang Rong$^*$ \quad Ahmet Enis Cetin$^*$\thanks{This work was supported by NSF IDEAL 2217023.}}
\address{$^*$Department of Electrical and Computer Engineering, University of Illinois Chicago, USA\\
$^\dagger$Machine \& Hybrid Intelligence Lab, Northwestern University, USA}
\begin{document}

%
\maketitle\small

\begin{abstract}
Electroencephalogram (EEG) data compression is necessary for wireless recording applications to reduce the amount of data that needs to be transmitted. In this paper, an asymmetrical sparse autoencoder with a discrete cosine transform (DCT) layer is proposed to compress EEG signals. The encoder module of the autoencoder has a combination of a fully connected linear layer and the DCT layer to reduce redundant data using hard-thresholding nonlinearity. Furthermore, the DCT layer includes trainable hard-thresholding parameters and scaling layers to give emphasis or de-emphasis on individual DCT coefficients. Finally, the one-by-one convolutional layer generates the latent space.  The sparsity penalty-based cost function is employed to keep the feature map as sparse as possible in the latent space. The latent space data is transmitted to the receiver. The decoder module of the autoencoder is designed using the inverse DCT and two fully connected linear layers to improve the accuracy of data reconstruction. In comparison to other state-of-the-art methods, the proposed method significantly improves the average quality score in various data compression experiments.

\end{abstract}
\section{Introduction}
\label{sec:intro}

Electroencephalogram (EEG) plays a crucial role in neurological diagnosis, including epileptic illness diagnosis, brain inflammation, and dementia \cite{al2022novel}. In existing literature, advanced feature extraction and classification algorithms have been designed for EEG analysis. Most of these algorithms gain insights into disease diagnosis by leveraging the abundance of EEG data detected by the sensor, which requires compression algorithms with high efficiency to establish high-capacity storage, fast transmission, and real-time analysis~\cite{cetin2006compression}. 

The current methods for compressing EEG signals are primarily categorized into three groups: traditional signal transform methods, neural network-based methods, and transform-based learning techniques.
In ~\cite{hussein2015scalable}, the discrete wavelet transform is applied to compress EEG data. It employed an optimization strategy to calculate the optimal control parameters which minimize the distortion and keep power consumption under a determined threshold. The work in ~\cite{nguyen2018study} developed a lossy compression model by utilizing the characteristics of epileptic EEG signals based on adaptive arithmetic encoding and discrete wavelet transform. However, these transform methods can not achieve a high reconstruction accuracy.

Neural network-based methods provide an alternative data compression approach by training models to learn the underlying patterns in the EEG signals. 
In ~\cite{al2018convolutional}, a convolutional autoencoder (CAE) is proposed to compress EEG data by employing convolutional layers and max-pooling layers to reduce the redundant data. Additionally, to retain more important information during the compression process, a new convolutional autoencoder is designed using e dynamic time warping (DTW) approximation as a loss function~\cite{lerogeron2023learning}.   

Transform-based learning methods have attracted considerable attention in the field of EEG compression over the past few decades \cite{cetin2006compression}. A near-lossless compression algorithm is developed based on discrete cosine transform (DCT) and multilayer perceptron (MLP) in ~\cite{hejrati2017new}. The energy concentration properties of DCT are leveraged to effectively reduce redundant data, while the MLP is utilized to compress the main DCT coefficients. Additionally, this approach calculates the reconstruction error to improve the accuracy. 
However, it does not have a good generalization ability
in the transfer learning experiment.

To overcome the limitations of low reconstruction accuracy in transform-based methods and low compression efficiency in neural network-based methods, this paper proposes an asymmetrical sparse autoencoder with a DCT layer for EEG sensor data compression. We introduced the DCT and Hadamard transform domain layers into neural networks in a number of applications \cite{pan2022deep,badawi2021discrete,badawi2022deep}. In this paper, the key idea is to perform elementwise multiplications in the transform domain as convolutions in the time domain and use the well-known soft and hard-thresholding units \cite{donoho1995noising} as the key nonlinearity of the network instead of the RELU. In EEG data compression trainable thresholding units not only remove the noise in the data but also improve the data compression efficiency. 
A single fully connected or convolutional layer cannot be trained together with a soft-thresholding or hard-thresholding nonlinearity. However, by adding the fixed DCT after a fully connected layer we can not only train the thresholds but also train the fully connected layer which adapts the data and improves the data compaction capability 
of DCT.
Furthermore, the  DCT layer has trainable scaling parameters and a one-by-one convolutional layer. Scaling parameters approximately perform filtering in the transform domain and the one-by-one convolution layer reduces the dimension and produces the latent space where the EEG data is compressed.
The decoder module reconstructs the EEG waveform from compressed data by using Inverse DCT (IDCT) and two fully connected linear layers because the EEG decoder systems have more computational power compared to the encoder module. 

The proposed model achieves the best compression efficiency and reconstruction accuracy compared with other autoencoder-based models. Experimental results show the proposed model outperforms other models on two EEG datasets: the BCI2~\cite{blankertz2001classifying} and the Bonn University~\cite{andrzejak2001indications} datasets in terms of quality score.
 Additionally, the proposed model has low computation cost at the encoder side, which is suitable for implementation at the edge in sensors.

\begin{figure*}[htbp]
	\centering
		\includegraphics[scale=.10]{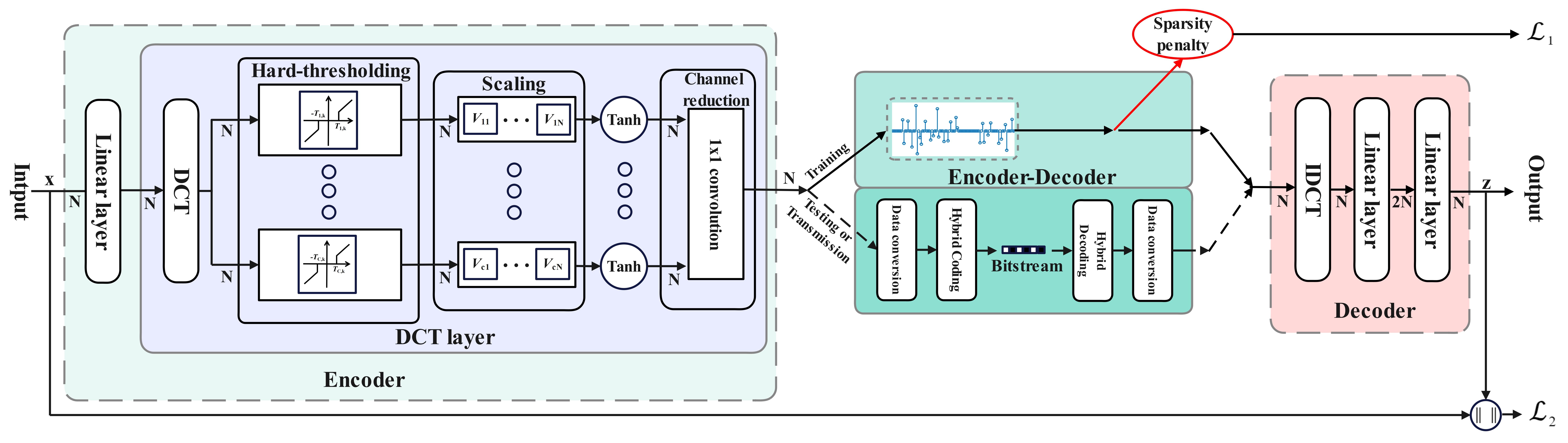}
	\caption{Block diagram of the Asymmetrical sparse autoencoder with a DCT layer.}
	\label{Fig:ASAEDCT}
\end{figure*}

\section{Preliminaries}
\label{sec:Background}
Our autoencoder structure has a DCT layer as shown in Fig.~\ref{Fig:ASAEDCT}.
The DCT \cite{ahmed1974discrete} is a real-valued transform, which utilizes cosine functions as its basis functions. 
In this work, type-III DCT is employed because of its convolution property. 
Given an input vector $\mathbf{x}=[x_0,x_1,\dots,x_{N-1}]$, its orthogonal type-III DCT $\mathbf{X}=[X_0,X_1,\dots,X_{N-1}]$ is defined as:
\begin{equation}
    {X}_k=\sqrt{\frac{1}{N}}x_0 +\sqrt{\frac{2}{N}}\sum_{n=1}^{N-1}x_n\text{cos}\left[\frac{\pi}{N}\left(k+\frac{1}{2}\right)n\right],
\label{Eq:dct3}
\end{equation}
Because of its good energy concentration properties, it has been widely used in many prevalent compression algorithms, including 
JPEG and MPEG algorithms \cite{wallace1992jpeg,sripada2006mp3}, where the DCT coefficients are weighted according to their importance before quantization. The weights are experimentally determined in these old standards. In this article, we will determine them using the backpropagation algorithm and call them scaling parameters.  In fact, applying weights to DCT parameters is somewhat similar to the performing convolution in the Fourier transform because elementwise multiplication in the Fourier transform corresponds to circular convolution in the time domain~\cite{xu2021dct}.
The DCT convolution theorem also states the element-wise multiplication in the DCT domain corresponding to symmetric convolution in the time domain~\cite{martucci1994symmetric}. Formally, it can be written as:
\begin{equation}\label{Eq:normal_dct_theorem}
    \mathbf{x} \otimes_s \mathbf{w} = {\mathscr{D}}^{-1}({\mathscr{D}}(\mathbf{x}\circ\mathbf{f}) \circ {\mathscr{D}}(\mathbf{w}\circ\mathbf{f}))\circ\mathbf{g},
\end{equation}
where $\mathbf{x}\in\mathbb{R}^N$ and $\mathbf{w}\in\mathbb{R}^N$ are input vectors.
$\circ$ represents the element-wise multiplication. $\otimes_s$ denotes the symmetric convolution~\cite{park2003m}. $\mathscr{D}(\cdot)$ and $\mathscr{D}^{-1}(\cdot)$ stands for the orthogonal type-III DCT and IDCT, respectively. $\mathbf{g}$ is a constant vector:
\begin{equation}\label{Eq:G}
\mathbf{g}[n]=\left\{
\begin{aligned}
&1/(2\sqrt{{N}}),  &n=0,\\
&\sqrt{1/(2N)} , &n > 0,\\
\end{aligned}
\right.
\end{equation}
where $0\leq n\leq N-1$. $\mathbf{f}[n]={1}/{\mathbf{g}[n]}$.

We experimentally observed that we do not need to perform symmetric convolutions to take advantage of the elementwise multiplications in the transform domain. We were able to train both the DCT domain weights and thresholding parameters using the backpropagation algorithm by inserting a fully connected layer before the DCT. We also observed that hard-thresholding performs better than soft-thresholding in EEG data compression.

\section{Asymmetrical Sparse Autoencoder with a DCT Layer}
In this section, we first describe the main building blocks of the proposed asymmetrical sparse autoencoder with a DCT layer (ASAEDCT). Then we explain its training using a cost function introducing sparsity.

Classical autoencoders are designed in a symmetric architecture, with a similar complexity on both the encoder and decoder sides~\cite{dinashi2022compression}. The encoder side needs to perform heavy computations to compress the signals. The computational load from the encoder prevents the application of the autoencoders in low-cost sensors. In ASAEDCT, the encoder side has low complexity, but the decoder side has more linear layers than the encoder side to enhance the data reconstruction ability because the decoder can be implemented in a powerful host computer. 

\noindent{\em Encoder Part of the ASAEDCT}:

The original EEG signal is partitioned into short-time windows of length $N$.
This block of data is processed by the fully connected linear layer. Let 
 the output of the linear layer be $\mathbf{x}\in\mathbb{R}^N$. The  DCT $\mathbf{X}= [X_{0}\ X_{1}\ \ldots\ X_{N}]^{\mathrm{T}}$ is computed using Eq.~(\ref{Eq:dct3}).

We have multiple channels to process the DCT data. Each channel corresponds to a different convolutional filter and each channel has
 hard-thresholding operators to perform denoising in the DCT domain. 
The proposed method applies the back-propagation algorithm to train hard-thresholding parameters and the transform domain weights. 
The hard-thresholding operator is defined as:
\begin{equation}
    \begin{split}
        \widetilde{X}_{i,k}= \mathcal{S_{T_{\rm{i}}}}\left(X_k\right)+T_{i,k}\cdot\text{sign}\left(\mathcal{S_{T_{\rm{i}}}}\left(X_{k}\right)\right),
    \end{split}
    \label{Eq:HT}
\end{equation}
where 
\begin{equation}
    \mathcal{S_{T_{\rm{i}}}}\left(X_k\right)= \text{sign}\left(X_k\right)\cdot\left(|X_k|-T_{i,k}\right)_{+}
\end{equation}
is the soft-thresholding function~\cite{donoho1995noising} and
$T_{i,k}$ is a trainable threshold parameter for $0\leq i \leq C$, $0\leq k \leq N$; the subscript
$C$ stands for the number of channels; $(\cdot)_+$ is the rectified linear unit (ReLU) function.
Although the soft-thresholding function can also perform transform domain denoising, it may also reduce the energy of large DCT coefficients. In EEG compression we found out that hard-thresholding produces better results.

 Unlike the hand-crafted quantization matrix used in the
 JPEG and MPEG-type standards~\cite{wallace1992jpeg}, our network has trainable scaling vectors to assign suitable weights for different DCT coefficients. 
  The DCT coefficients $\widetilde{X}_{i,k}$ are element-wise multiplied by the scaling parameters:
\begin{eqnarray}\label{Eq:scaling}
\widehat{X}_{i,k}=\widetilde{X}_{i,k}\cdot V_{i,k},
\end{eqnarray}
for $0 \leq k \leq N-1$. Another interpretation of the scaling layer is related to the DCT convolution theorem, i.e.,
$C$ scaling channels correspond to $C$ distinct convolution kernels in the time domain,
and their combination forms a filter bank to enhance the ability to extract features.

The one-by-one convolutional layer~\cite{lin2013network} is used to reduce the number of channels and forms the latent space where the EEG data is compressed. 
After the scaling layer, tanh
is employed to introduce a nonlinearity between the scaling layer and the one-by-one convolution layer. 
The overall operation in the DCT layer can be summarized in the Algorithm~\ref{algorithm:DCT}.
\begin{algorithm}[htbp]
 \caption{C-path DCT layer}
 \label{algorithm:DCT}
  \KwIn{Input tensor ${\mathbf{{x}}} \in\mathbb{R}^{N}$}
  \KwOut{Output tensor ${\mathbf{{y}}} \in\mathbb{R}^{N}$}
  \bf{Define:} $\mathbf{T}_i\in\mathbb{R}^{N}, \mathbf{V}_i\in\mathbb{R}^{N}, \mathbf{F}=\rm{Conv1D}(in=C, out=1,~kernel~size=1)$;\\
        ${\mathbf{X}}={\mathbf{DCT}}(\mathbf{x})\in\mathbb{R}^{N};$ \\
        \For{$i=1;i \le C;i=i+1$}
        {
            $\mathbf{U}_i= {\rm{sign}}\left(\mathbf{X}_i\right)\left(|\mathbf{X}_i|-\mathbf{T}_i\right)_{+}\in\mathbb{R}^{N};$\\
            $\mathbf{\widetilde{X}}_i= \mathbf{U}_i+\mathbf{T}_i\circ{\rm{sign}}\left(\mathbf{U}_i\right)\in\mathbb{R}^{N};$\\
            $\mathbf{\widehat{X}}_i=\mathbf{\widetilde{X}}_i\circ\mathbf{V}_i\in\mathbb{R}^{N};$\\
            $\mathbf{R}_i={\rm{tanh}}(\mathbf{\widehat{X}}_i)\in\mathbb{R}^{N};$ \\
        }
        $\mathbf{R}={\rm{stack}}(\mathbf{R}_i)\in\mathbb{R}^{C\times N};$ \\
        $\mathbf{y}={\rm{F}}(\mathbf{R})\in\mathbb{R}^{N};$ \\
        return ${\mathbf{y}}\in\mathbb{R}^{N};$ \\
\end{algorithm}

We observed that $C=3$ produces the best results in EEG data compression (to be discussed in Section~\ref{Experimental Results}). Another observation that we had is that we obtained better coding results when we changed the order of scaling the hard-thresholding nonlinearity in our network as shown in Fig.1.

\noindent{\em Decoder Part of the ASAEDCT:}\\
Encoder output data is in the DCT domain and it is quantized as discussed in Subsection~\ref{sec:transmission}. 
Once the decoder receives the quantized data $\widehat{\mathbf{X}}$ in the DCT domain, the decoder first computes its IDCT $\hat{\mathbf{x}}$. After this stage,  the output of the network is obtained using two fully connected layers as shown in Fig.~\ref{Fig:ASAEDCT}. Compared with the encoder, more linear fully connected layers are included in the decoder to enhance the reconstruction ability.


\subsubsection{Sparsity Penalty Based Cost Function}
\label{sec:Sparsity Penalty}

In this subsection, we describe the cost function used in training the proposed ASAEDCT system.  Sparsity constraints~\cite{ng2011sparse} are introduced to keep the feature map sparse, which enhances the compression efficiency of the DCT layer. Suppose the output of the one-by-one convolutional layer is $\mathbf{y}\in \mathbb{R}^N$, the activity of ${y}_{j}$ is defined as:
\begin{equation}
    \hat{y}_{j}={\sigma}(\mathbf{y})_{j}=\frac{e^{\lvert y_{j}\rvert}}{\sum_{k=0}^{N-1}e^{\lvert y_{k}\rvert}}, i=0,1,\cdots,N-1,
\end{equation}
where $\sigma(\cdot)$ stands for the softmax function.
Next, the Kullback–Leibler divergence (KLD)~\cite{ng2011sparse} is utilized as the sparsity penalty term because it can measure the difference between two probability distributions. The KLD is defined as
\begin{equation}
\sum_{j=0}^{N-1} {\rm{KL}}(\alpha\vert\vert\hat{y}_{j})
=\sum_{j=0}^{N-1}\alpha\log\frac{\alpha}{\hat{y}_{j}}+
    (1-\alpha)\log\frac{1-\alpha}{1-\hat{y}_{j}},
    \label{Eq:KL}
\end{equation}
where $\alpha$ is a sparsity parameter. 
Therefore, the overall loss function $\mathcal{L}$ is composed of the linear combination of the mean squared error and the KLD:
\begin{align}
&\mathcal{L}=\frac{1}{N}\sum_{i=0}^{N-1} \left({x}_i-z_i\right)^2+\lambda\sum_{j=0}^{N-1} {\rm{KL}}\left(\alpha\vert\vert{\sigma}(\mathbf{y})_j\right),
    \label{Eq:loss_p}
\end{align}
where $\lambda$ is the weight of the sparsity penalty term. As shown in Fig~\ref{Fig:ASAEDCT}, $x_i$ and  $z_i$ represent the input and reconstructed signals, respectively. In the training phase, a threshold $\xi$ is employed to adjust the compression ratio. When the proportion of 0's in $\mathbf{y}$ is lower than $\xi$, the training is terminated. In this manner, small entries (redundant information) are eliminated by being set to 0. The setting of $\xi$ is presented in Section~\ref{Experimental Results}.

\subsection{Data Encoding and Storage}
\label{sec:transmission}
As is shown in Fig.~\ref{Fig:ASAEDCT}, the data transmission module consists of data conversion and hybrid coding. The hybrid coding algorithm is developed using a combination of Run-Length Encoding (RLE) \cite{akhter2010ecg} and Lempel–Ziv–Markov chain algorithm (LZMA) \cite{tu2006novel}.
Each double floating-point data requires a storage space of 64 bits in the memory, whereas an integer only needs 32 bits~\cite{ratanaworabhan2006fast}. Therefore, floating-point numbers are converted to integers. The data conversion is formulated as:
\begin{eqnarray}\label{Eq:code}
\widehat{\mathbf{y}}=\text{Round}(10^{\theta}\times {\mathbf{y}}/\phi) ,    
\end{eqnarray}
where $\text{Round}(\cdot)$ is the integer rounding function; $\theta$ and $\phi$ are both integers. They are used to adjust the compression ratio. 
This data is encoded into a bitstream for wireless transmission or storage into an ambulatory device using the RLE and LZMA. The RLE is first employed to eliminate consecutive repeated zeros in the latent space. Then, LZMA is implemented on the output of the RLE.
At the decoder, this bitstream is converted back into integers and processed by the decoder part of ASAEDCT for signal reconstruction in the host computer.  

\section{Experimental Results}
\label{Experimental Results}
In this work, the BCI2 dataset~\cite{blankertz2001classifying}, and the Bonn
University dataset~\cite{andrzejak2001indications} are used to evaluate the performance of the proposed EEG compression scheme.
(1) The BCI2 dataset~\cite{blankertz2001classifying} is collected from a normal subject using 28 EEG channels. The sampling frequency is 100Hz. It contains 316 training sets and 100 testing sets. Each channel includes 50 samples. (2) The Bonn University dataset~\cite{andrzejak2001indications} includes five records, \textit{i.e.}, F, N, O, Z, S. To evaluate the generalization ability of the model, we use record S which is collected from an epileptic subject. All data sets are partitioned into blocks with a size of $N=64$. The AdamW optimizer~\cite{loshchilov2017decoupled} is utilized in the training process. We choose $\theta$, $\phi$ and $\xi$ as 4, 5 and 0.6 because it can achieve a balance between compression efficiency and reconstruction accuracy.
Additionally, the batch size is 16, and the learning rate is 0.001. $\lambda$ is 10. Moreover, we use the compression ratio (CR)~\cite{pal2023optimized} and percent root-mean-square difference (PRD)~\cite{jha2018electrocardiogram} as the performance metrics. The comparison efficiency ascends as the CR increases, and the difference between the constructed signal and the original signal reduces as the PRD decreases. 

To validate the compression performance of the proposed model, we compare our model with the ANN+DCT~\cite{hejrati2017new} which achieved the best result on the BCI2 and Bonn datasets so far. Additionally, we consider the transform-based method~\cite{jha2018electrocardiogram}, the neural networks-based method~\cite{ng2011sparse} and the transform-based learning method~\cite{pan2023real} for further comparison as they both have a low complexity encoder.

Table~\ref{tab: experimental results} provides the detailed data compression performance of the proposed methods versus the other four state-of-the-art algorithms on the BCI2 datasets. In comparison to sparse AE, the proposed approach increases CR from 21.38 to 25.66 and decreases PRD from 8.51 to 5.09. Additionally, our model is superior to  DCT-DOST because it has a higher CR and a lower PRD. It indicates that ASAEDCT which combines the orthogonal DCT transform and a neural network can achieve a better compression performance. Compared with other transform-based learning compression methods~\cite{hejrati2017new, pan2023real}, ASAEDCT still provides a better quality score. This is because the DCT layer improves compression efficiency and asymmetrical structure enhances the reconstruction ability in the decoder. Moreover, to validate the robustness of the model, the model trained using the BCI2 dataset is also tested on the Bonn University dataset. It is observed that the AAESDCT with $C=3$ channels outperforms other methods as it achieves the best QS. Therefore, the model has a good generalization capability in data compression tasks.

\begin{table}[ht]
\centering
\begin{tabular}{|c|c|c|c|c|}
\hline
\makecell{\bf{Test Data}}&\makecell{\bf{Algorithm}}&\makecell{\bf{CR}} &\makecell{\bf{PRD {(\%)}}}& \makecell{\bf{QS}} \\
\hline   
 \multirow{5}{*}{BCI2}
 &AAN+DCT~\cite{hejrati2017new}    &\makecell{21.57} & \makecell{6.38} & \makecell{3.38} \\
 &Sparse AE~\cite{ng2011sparse}    &\makecell{21.38} & \makecell{8.51} & \makecell{2.51} \\
 &DCT-DOST~\cite{jha2018electrocardiogram}    &\makecell{21.67} & \makecell{10.40} & \makecell{2.08} \\
 &AE-DCT~\cite{pan2023real}   &\makecell{21.95} & \makecell{9.18} & \makecell{2.39} \\
 &\bf{ASAEDCT}   &\bf{\makecell{25.66}} & \bf{\makecell{5.09}} & \bf{\makecell{5.05}} \\
\hline   
\multirow{5}{*}{Bonn}
 &AAN+DCT~\cite{hejrati2017new}    &\makecell{4.45} & \makecell{7.22} & \makecell{0.62} \\
 &Sparse AE~\cite{ng2011sparse}    &\makecell{6.76} & \makecell{6.18} & \makecell{1.09} \\
 &DCT-DOST~\cite{jha2018electrocardiogram}    &\makecell{7.56} & \makecell{11.09} & \makecell{0.68} \\
 &AE-DCT~\cite{pan2023real}   &\makecell{6.84} & \makecell{6.65} & \makecell{1.03}   \\
 &\bf{ASAEDCT}   &\bf{\makecell{8.40}} & \bf{\makecell{5.89}} & \bf{\makecell{1.42}} \\
\hline
\end{tabular}
\caption{Compression experimental results. All methods are trained on the BCI2 dataset. }
\label{tab: experimental results}
\end{table}

\begin{figure}[htbp]
	\centering
		\includegraphics[scale=.38]{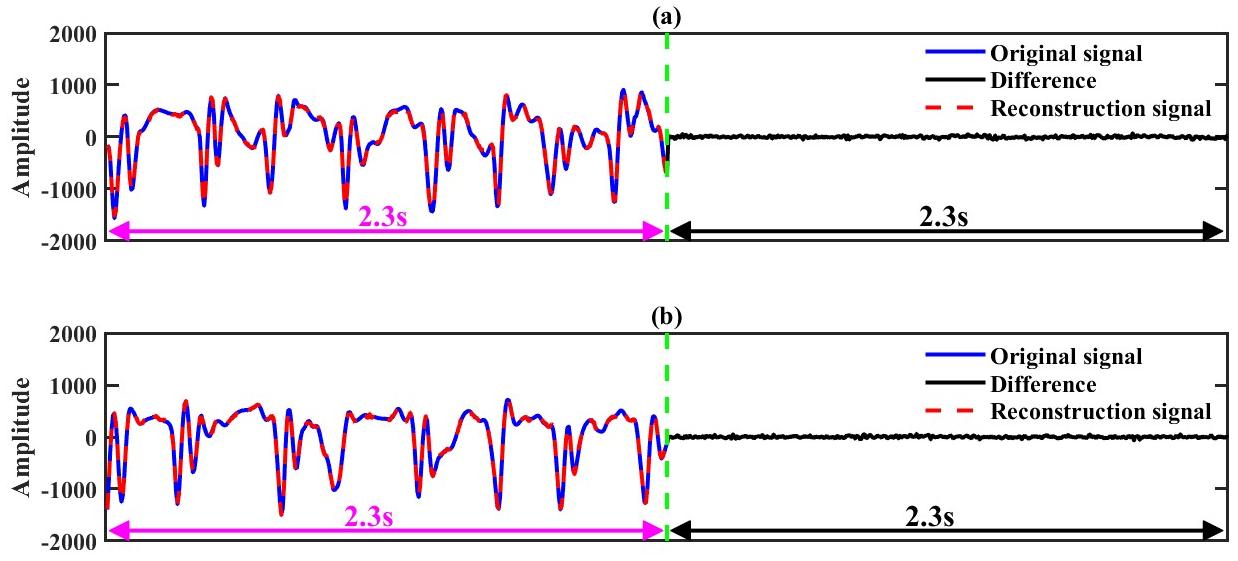}
	\caption{Comparison of the time domain between the original data and the reconstructed data on the Bonn dataset.}
	\label{SEUF}
\end{figure}

For visual evaluation of the compression performance of the proposed methods, the original signals and reconstruction signals are depicted in Fig~\ref{SEUF}. It is observed that the original signals and reconstruction signals are similar to each other as the differences between them are limited to a small range. 

\begin{table}[htbp]
\centering
\begin{tabular}{|p{3cm}|p{1.0cm} p{1.4cm} p{1.0cm}|}
\hline
\makecell{\bf{Algorithm}}&\makecell{\bf{CR}} &\makecell{\bf{PRD {(\%)}}}& \makecell{\bf{QS}} \\
\hline   
 No nonlinearity &\makecell{25.12} & \makecell{5.17} & \makecell{4.86}  \\
 No scaling    &\makecell{25.46} & \makecell{11.79} & \makecell{2.16}  \\
 No penalty   &\makecell{24.51} & \makecell{5.26} & \makecell{4.66}  \\
 Only DCT  &\makecell{25.44} & \makecell{7.05} & \makecell{3.61} \\
 With soft-threshold  &\makecell{25.34} & \makecell{11.62} & \makecell{2.18} \\
 One-channel (C=1)  &\makecell{25.07} & \makecell{6.00} & \makecell{4.18}   \\
 Two-channel (C=2) &\makecell{24.90} & \makecell{5.39} & \makecell{4.62}   \\
 \bf{ASAEDCT}(C=3)  &\bf{\makecell{25.66}} & \bf{\makecell{5.09}} & \bf{\makecell{5.05}} \\
 Four-channel (C=4)  &\makecell{25.01} & \makecell{5.31} & \makecell{4.71}   \\
\hline
\end{tabular}
\caption{Ablation experimental result. }
\label{tab3: experimental results}
\end{table}

 Table~\ref{tab3: experimental results} presents the ablation study of each module in ASAEDCT. When the scaling layer, sparsity penalty term, or nonlinearity are removed, the CR decreases and PRD increases, leading to a lower QS. Additionally, the three-channel $(C=3)$ model is superior to one-channel, two-channel, and four-channel models. When we use only the DCT to compress the EEG data, the QS decreases. When hard thresholds are replaced with soft thresholds, the performance degrades.

Lastly, an analysis of the computational cost related to the proposed method is conducted.
For an input with a duration of 6.4s, the compression time of the ASAEDCT network is only around 0.016s.  Although this experiment is executed on the Intel Core i7-12700H CPU, the fast compression time indicates that the proposed method can be used in real-time compression of the EEG signals. 
\section{Conclusion}
This study presents an asymmetrical autoencoder with a DCT layer that contains hard-thresholding nonlinearity for EEG sensor data compression. Since the encoder module uses the combination of one fully connected linear layer and the DCT layer the threshold values and DCT domain weights can be trained using a backpropagation type algorithm. The hard-thresholding nonlinearity and scaling layers not only enhance the data compaction capability of the DCT layer but also denoise the EEG signal. The decoder module based on the IDCT and two linear layers reconstructs the original signal.  The proposed model achieves the best compression efficiency and reconstruction accuracy compared to other methods in BCI2 and Bonn EEG datasets. The computational load of the compression part of the ASAEDCT network is low, therefore, it can be implemented in the sensor for EEG data compression.
\small
\bibliographystyle{unsrt}
\bibliography{ref}

\end{document}